\documentclass[12pt,fleqn]{article}

\usepackage{epsfig}
\usepackage{amsmath}
\usepackage{times}
\usepackage[sort&compress,comma]{natbib}
\usepackage{amssymb}

\title{\sffamily\textbf{Continuously stable strategies as evolutionary branching points}}
\author{Michael Doebeli \& Iaroslav Ispolatov\\\\
\vspace{-2mm}\normalsize Departments of Zoology and Mathematics\\
\vspace{-2mm}\normalsize University of British Columbia, 6270 University Boulevard\\
\vspace{-2mm}\normalsize Vancouver B.C. Canada, V6T 1Z4}
\date{\normalsize\today}

\begin{document}
\maketitle

%\paragraph{Keywords:} evolutionary game theory; cooperation; prisoner's dilemma; population dynamics

%\paragraph{Word count:} abstract: XXX; body: XXX; figure legends: XXX; 3 figures.

%\subsubsection*{Corresponding author:}
%who?

%\newpage

\begin{center} {\bf \large Abstract}
\end{center}
\noindent
Evolutionary branching points are a paradigmatic feature of adaptive dynamics, because they are potential starting points for adaptive diversification. The antithesis to evolutionary branching points are Continuously stable strategies (CSS's), which are convergent stable and evolutionarily stable equilibrium points of the adaptive dynamics and hence are thought to represent endpoints of adaptive processes. However, this assessment is based on situations in which the invasion fitness function determining the adaptive dynamics have non-zero second derivatives at a CSS. Here we show that the scope of evolutionary branching can increase if the invasion fitness function vanishes to higher than first order at a CSS. Using a class of classical models for frequency-dependent competition, we show that if the invasion fitness vanishes to higher orders, a CSS may be the starting point for evolutionary branching, with the only additional requirement that mutant types need to reach a certain threshold frequency, which can happen e.g. due to demographic stochasticity. Thus, when invasion fitness functions vanish to higher than first order at equilibrium points of the adaptive dynamics, evolutionary diversification can occur even after convergence to an evolutionarily stable strategy.

\vskip 2cm

{\bf \large Introduction}
\vskip 1cm
Since the publication of the seminal papers \cite{metz_etal1992}, \cite{metz_etal1996}, \cite{dieckmann_law1996} and \cite{geritz_etal1998}, the theory of adaptive dynamics has attracted a lot of attention as a useful tool to study long-term evolutionary dynamics in continuous phenotype spaces. In particular, the phenomenon of evolutionary branching emerged as paradigmatic feature of adaptive dynamics that serves as a basic metaphor for adaptive diversification and speciation \citep{geritz_etal1998, dieckmann_doebeli1999, geritz_kisdi2000, doebeli_dieckmann2003, dieckmann_etal2004, doebeli2010}. During evolutionary branching, a population first evolves under directional selection to an evolutionary branching point in phenotype space, at which selection turns disruptive due to frequency dependence, which in turn causes the population to split into diverging phenotypic clusters. 

Evolutionary branching is a generic feature of adaptive dynamics and can be observed in a multitude of models incorporating frequency-dependent selection \citep{doebeli2010}. The analytical understanding of evolutionary branching rests on the distinction between two basic stability concepts of adaptive dynamics, which are easiest to understand in the context of adaptive dynamics in 1-dimensional trait spaces. In general, adaptive dynamics is derived from the invasion fitness function function  $f(x,y)$ describing the long-term per capita growth rate of a rare mutant type $y$ in a resident population that is monomorphic for trait value $x$ \citep{metz_etal1996, geritz_etal1998}. The adaptive dynamics of the trait $x$ is then given as a gradient dynamics of the invasion fitness function and described by the canonical equation \citep{dieckmann_law1996}:
\begin{align}
\frac{dx}{dt}=M(x)\left.\frac{\partial f(x,y)}{\partial y}\right\vert_{y=x}.
\end{align}
The partial derivative with respect to the mutant trait is evaluated at the resident trait, and the selection gradient $\partial f/\partial y$ is in general a vector of the same dimension as the phenotype space. $M(x)$ is a matrix describing the mutational process in the phenotype under consideration, but for 1-dimensional trait spaces, $M(x)$ is a scalar that only influences the speed of evolution. For 1-dimensional traits one therefore often assumes $M(x)=1$ to simplify the analysis. Equilibrium points of the adaptive dynamics are then given as points $x^*$ in phenotype space at which the selection gradient vanishes, i.e., points $x^*$ satisfying
\begin{align}
\left.\frac{\partial f(x^*,y)}{\partial y}\right\vert_{y=x^*}=0.
\end{align}
A singular point is called convergence stable if it is locally stable for the adaptive dynamics (1), i.e., if the system (1) returns to the singular point after any perturbation that is small enough. On the other hand, a singular point is called evolutionarily stable
if the invasion fitness function $f(x^*,y)$ has, as a function of $y$, a maximum at $x^*$, for in that case no mutant trait in the vicinity of the singular point has a higher growth rate than the singular trait value itself.

Even though both types of stability, convergence stability and evolutionary stability, are derived from the invasion fitness function $f(x,y)$, the two conditions for stability are in general different. For example, it is possible that a singular point is evolutionarily stable, but not convergence stable, a situation that is referred to as Garden-of-Eden configuration \citep{nowak_sigmund1990}, because even though a resident with the singular trait value cannot be invaded by any nearby mutant, the resident population will actually evolve away from the singular point if it is started with a resident value that is arbitrarily close (but not equal) to the singular point. 

On the other hand, it is also possible that a singular point is convergent stable, but not evolutionarily stable, in which case the population will evolve to the singular point, only to find itself being vulnerable to invasion by any nearby mutant. In such a situation the singular point is a minimum for the invasion fitness function, and convergence to such a point therefore opens the scope for evolutionary diversification. 

When the invasion fitness function only vanishes to first order at a singular point (i.e., only the selection gradient, but not the higher order derivatives of the invasion fitness vanish at the singular point), the adaptive dynamics around a singular point can be fully classified in terms of the second derivative of the invasion fitness at the singular point \citep{geritz_etal1998, diekmann2003, doebeli2010}. In short, a convergence stable singular point (and only those are really of interest) is either an endpoint of the evolutionary dynamics, or it is an evolutionary branching point at which an evolving population diversifies into diverging phenotypic clusters. In the first case, the convergent stable singular point is also evolutionarily stable, so that after convergence to the singular point, the population is immune to invasion by nearby mutants. Such points are classically called Continuously Stable Strategies (CSS; \cite{eshel1981}) and represent evolutionary endpoints at least under the assumption of small mutations.

In the second case, the convergent stable singular point is evolutionarily unstable, and it can be shown that in 1-dimensional trait spaces, these two conditions imply two more conditions: that of mutual invasibility, and that of evolutionary divergence. Mutual invasibility refers to the fact that trait values on either side of the singular can mutually invade each other. More precisely, if $\epsilon>$ is small enough, then a resident population that is monomorphic for trait value $x^*-\epsilon$ can be invaded by a mutant $x^*+\epsilon$, and vice versa. In particular, phenotypes on either side of the singular point can coexist in a protected polymorphism. Moreover, in such a polymorphic population consisting of traits $x^*-\epsilon$ and $x^*+\epsilon$, evolution in each monomorphic subpopulation occurs away from the singular point, so that the phenotypic distance between the two coexisting clusters increases, and hence the clusters diverge evolutionarily.

When the invasion fitness function only vanishes to first order at a singular point, a CSS may satisfy the condition for mutual invasibility, but it never satisfies the condition for evolutionary divergence of coexisting subpopulations. Thus, while phenotypes on either side of a CSS may coexist, evolution in two coexisting phenotypic clusters always induces convergence of both clusters to the singular trait value. In particular, a CSS cannot be the starting point for evolutionary diversification.  

These statements can be made precise in terms of the second derivatives of the invasion fitness function at the singular point \citep{geritz_etal1998, diekmann2003}. In particular, the condition for convergence stability of the singular point, i.e., for local stability of an equilibrium of the dynamical system (1), is given by
\begin{align}
\frac{d}{dx}\left[\left.\frac{\partial f(x,y)}{\partial y}\right\vert_{y=x=x^*}\right]=\left.\frac{\partial^2 f(x,y)}{\partial x\partial y}\right\vert_{y=x=x^*}+\left.\frac{\partial^2 f(x,y)}{\partial y^2}\right\vert_{y=x=x^*}<0.
\end{align}
Also, the condition for evolutionary stability of a singular point is simply
\begin{align}
\left.\frac{\partial^2 f(x,y)}{\partial y^2}\right\vert_{y=x=x^*}<0.
\end{align}
%%%%%%%%%%%%%%%%%%%%%%%%%%%%%%%%%%%%%%%%%%%%%%%%%%%%%%%%%%%%%%%%
If both these conditions are satisfied, the singular point $x^*$ is a CSS, and hence no diversification occurs. On the other hand, if the first equality is satisfied but the second is reversed (so that the singular point is a minimum of the invasion fitness function), then not only is the singular point convergent stable, but both mutual invasibility and evolutionary divergence as described above are satisfied, and hence $x^*$ is an evolutionary branching point \citep{diekmann2003, doebeli2010}.

In this paper, we show that the scope of evolutionary branching can increase if the invasion fitness function vanishes to higher than first order at a singular point. More precisely, using a class of classical models for frequency-dependent competition we show that if the invasion fitness vanishes to higher orders, a CSS may satisfy both the conditions for mutual invasibility and the condition for evolutionary divergence, and hence can be the starting point for evolutionary branching. This can happen even if mutations are small, with the only additional requirement that mutant types need to reach a certain threshold frequency, e.g. due to demographic stochasticity. Thus, when invasion fitness functions vanish to higher than first order at equilibrium points of the adaptive dynamics, evolutionary diversification can occur even after convergence to an evolutionarily stable singular point.

\vskip 2 cm
{\bf \large Model and Results}
\vskip 1cm
We consider models for frequency-dependent competition in which individuals are characterized by a 1-dimensional phenotype $x$ (e.g., body size). The phenotype determines two ecological properties. On the one hand, the carrying capacity of populations that are monomorphic for trait value $x$ is given by the carrying capacity function $K(x)$, and on the other hand the strength of competition between two individuals with phenotypes $x$ and $y$ is given by the competition kernel $\alpha(x,y)$. Note that by definition, the carrying capacity is a property of populations, but the fact that $K(x)$ is the carrying capacity of monomorphic populations makes it a function of the phenotype $x$ (moreover, it is easy to define this function at the individual level in terms of per capita birth and death rates; see e.g. Chapter 3 in \cite{doebeli2010}).

The ecological dynamics of a resident population that is monomorphic for phenotype $x$ is assumed to be logistic:
\begin{align}
\frac{dN_x}{dt}=rN_x\left(1-\frac{N_x}{K(x)}\right),
\end{align} 
where $N_x$ is the population size of the resident $x$, and $r>0$ is the intrinsic growth rate, which is assumed to be independent of the phenotype $x$. Without loss of generality we set $r=1$ in the following. Clearly, $N_x=K(x)$ at ecological equilibrium. When a mutant $y$ appears in a resident $x$, the dynamics of its population density $N_y$ is given by 
\begin{align}
\frac{dN_y}{dt}=N_y\left(1-\frac{N_y+\alpha(y,x)N_x}{K(y)}\right),
\end{align} 
where the competition kernel $\alpha(y,x)$ is used to describe the competitive impact of the resident $x$ on the mutant $y$. If the resident is at its ecological equilibrium $K(x)$ and the mutant is rare, so that $N_y$ is negligible, the per capita growth rate of the mutant becomes
\begin{align}
f(x,y)=1-\frac{\alpha(y,x)K(x)}{K(y)}.
\end{align}
$f(x,y)$ is the invasion fitness function, which determines the adaptive dynamics.

Here we consider the following class of functions for the carrying capacity and the competition kernel:
\begin{align}
K(x)=&K_0\exp\left(-\frac{1}{2}\left(\frac{\vert x \vert }{\sigma_K}\right)^s\right)\\
\alpha(x,y)=&\exp\left(-\frac{1}{2}\left(\frac{\vert x-y\vert}{\sigma_\alpha}\right)^s\right)
\end{align}
We assume that the exponent is a real number $s\geq2$ to ensure that $K$ is at least twice differentiable at $0$ and $\alpha$ is at least twice differentiable at $y=x$. (Note that the first derivative of $K$ at 0 and of $\alpha$ at $y=x$ is 0.) The carrying $K(x)$ decreases with increasing distance from its unique maximum at $x=0$ at a rate that is determined by the parameter $\sigma_K$. In evolutionary terms, $K(x)$ describes a stabilizing component of selection whose strength is measured by $\sigma_K$ (with small $\sigma_K$ indicating strong stabilizing selection). The parameter $K_0$ scales the population density.

The competition kernel $\alpha(x,y)$ describes the frequency-dependent component of selection, and for any given trait value $x$, the strength of the competitive impact of individuals with trait value $y$ decreases with increasing distance $\vert y-x\vert$ at a rate that is determined by the parameter $\sigma_\alpha$, which therefore determines the strength of frequency dependence (with small $\sigma_\alpha$ indicating a rapid decline of competitive impacts with phenotypic distance). Note that the competition kernel is symmetric in the sense that it only depends on the phenotypic distance $\vert y-x\vert$. Also note that for $s=2$, the carrying capacity and the competition kernel have Gaussian form, which results in a classical model for frequency-dependent competition that has a long tradition in ecological and evolutionary theory \citep{dieckmann_doebeli1999, macarthur_levins1967,roughgarden1979,kirkpatrick_barton1997}.

With the invasion fitness function (7), the adaptive dynamics (1) becomes
\begin{align}
\nonumber\frac{dx}{dt}=&\left.\frac{\partial f(x,y)}{\partial y}\right\vert_{y=x}\\
=&-\left.\frac{\partial \alpha(y,x)}{\partial y}\right\vert_{y=x}+\frac{K'(x)}{K(x)}\\
\nonumber=&\frac{K'(x)}{K(x)}
\end{align}
(where for simplicity and without loss of generality we have assumed $M(x)=1$ for the mutational process). Expression (10) reflects the fact that away from singular points, the adaptive dynamics is solely determined by the stabilizing component of selection, i.e., by the carrying capacity function $K(x)$. This is essentially due to the assumption that the competition kernel is symmetric and differentiable at $y=x$, which implies that its derivative at $x=y$ must be 0. It immediately follows that $x^*=0$ is the only singular point of the adaptive dynamics, because $x^*=0$ is the only solution of $K'(x^*)=0$. Moreover, even though the second derivative of $K(x)$ at $x^*=0$ may be 0, and hence the Jacobian of the dynamical system (10) at the equilibrium $x^*=0$ may be 0, it is clear that $x^*=0$ is a globally stable equilibrium for the dynamics (10), because $dx/dt>0$ for $x<x^*$ and $dx/dt<0$ for $x>x^*$. Thus, $x^*=0$ is a convergent stable singular point.

Evolutionary stability of the singular point is determined  by the second derivative of the invasion fitness function. For $s=2$, we have 
\begin{align}
\left.\frac{\partial^2 f(0,y)}{\partial y^2}\right\vert_{y=0}=\frac{1}{\sigma_\alpha^2}-\frac{1}{\sigma_K^2}.
\end{align}
Therefore, $x^*=0$ is evolutionarily stable (i.e., the partial derivative (11) is negative), and hence a CSS, if and only if $\sigma_\alpha>\sigma_K$, which is a familiar result for the Gaussian case $s=2$ \citep{dieckmann_doebeli1999}. Also, $x^*$ is an evolutionary branching point, and hence the starting point of adaptive diversification, if and only if $\sigma_\alpha<\sigma_K$. The situation changes, however, for $s>2$.

For $s>2$, we have
\begin{align}
\left.\frac{\partial^2 f(0,y)}{\partial y^2}\right\vert_{y=0}=0,
\end{align}
but it is easy to see that, again, the invasion fitness $f(x^*,y)$ has a maximum at $y=0$ if and only if  $\sigma_\alpha>\sigma_K$. Thus, the evolutionary stability condition is the same for all $s\geq2$. In particular, the singular point is a CSS if and only if  $\sigma_\alpha>\sigma_K$. What changes, however, is that for $s>2$, the conditions for mutual invasibility and for evolutionary divergence that are necessary for evolutionary branching may be satisfied even if  $\sigma_\alpha>\sigma_K$, i.e., even if the singular point is a CSS.

To see this, we assume $\sigma_\alpha>\sigma_K$, and we first consider coexistence of two strategies $\epsilon>0$ and $-\epsilon$ close to the singular point $x^*=0$. (We note that we make the symmetry assumption of considering phenotypes $\epsilon$ and $-\epsilon$ for analytical convenience, and that the arguments given below are also valid for resident strains $\epsilon$ and $\delta$, as long as the two strains lie on either side of the singular point and $\vert \epsilon\vert$ and $\vert\delta\vert$ are small.) The per capita growth rate of a rare mutant $-\epsilon$ in a resident $\epsilon$ at equilibrium $K(\epsilon)$ is
\begin{align}
1-\frac{\alpha(-\epsilon,\epsilon)K(\epsilon)}{K(-\epsilon)}=1-\alpha(-\epsilon,\epsilon)>0,
\end{align}
and similarly for the growth rate of a rare $\epsilon$ mutant in a $-\epsilon$ resident. It follows that the two types can coexist, and if $\hat N_\epsilon$ denotes the equilibrium population density of the two subpopulations at the coexistence equilibrium (the two subpopulations have the same density at equilibrium due to symmetry), one easily calculates that 
\begin{align}
\hat N_\epsilon=\frac{K(\epsilon)(1-\alpha(\epsilon,-\epsilon))}{1-\alpha(\epsilon,-\epsilon)^2}
=\frac{K(\epsilon)}{1+\alpha(\epsilon,-\epsilon)}.
\end{align}
To check for evolutionary divergence, we assume two coexisting residents $\epsilon$ and $-\epsilon$ and calculate the invasion fitness $f(\epsilon,-\epsilon,y)$ for rare mutants $y$ occurring in either resident:
\begin{align}
f(\epsilon,-\epsilon,y)=1-\frac{N_\epsilon(\alpha(y,\epsilon)+\alpha(y,-\epsilon))}{K(y)}
\end{align}
A salient example of this invasion fitness function is shown in Figure 1, which illustrates that selection in the two resident strains $\epsilon$ and $-\epsilon$ points away from the singular point, i.e., that selection is divergent.

\begin{figure}[htp]
\centering
\epsfig{file=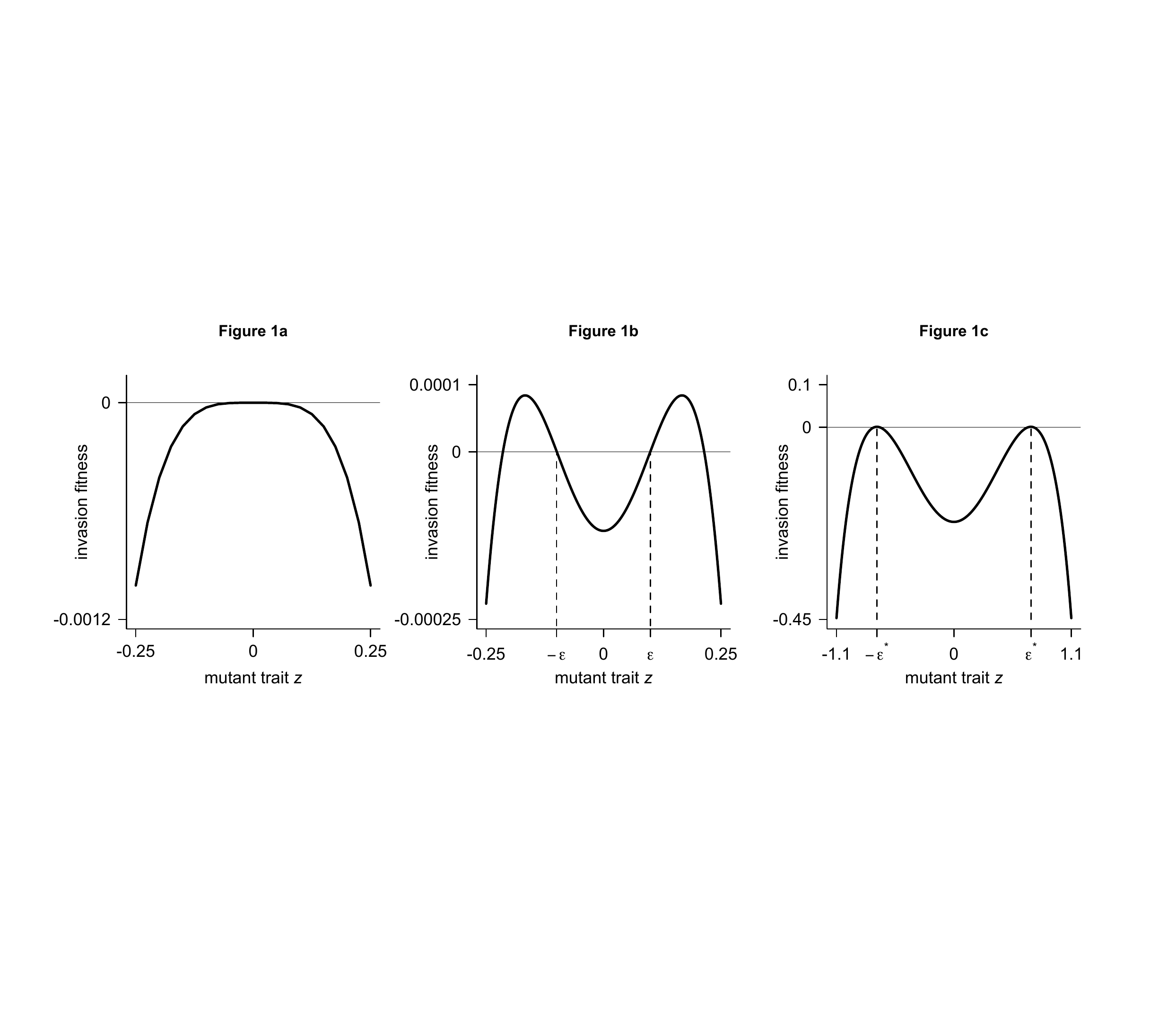,width=1\textwidth}
\caption{\label{Figure 1}
\footnotesize Invasion fitness as a function of the mutant trait $z$ when the resident is monomorphic for the singular strategy $x^*$ (panel a), and when the resident population consists of two coexisting strategies $\epsilon$ and $-\epsilon$ (panels b,c). Panel a shows that the singular point is a fitness maximum (and hence a CSS, see main text). Panel b shows that in the invasion fitness function with two residents $\epsilon$ and $-\epsilon$, the maximum at $x^*=0$ shown in panel a resolves into two maxima, one lying to the right of $\epsilon$ and the other lying to the left of $-\epsilon$. As a consequence, the derivative of the invasion fitness is positive at $\epsilon$ and negative at $-\epsilon$, leading to divergent evolution (note that the fitness function is necessarily 0 at the two resident strategies). Panel c illustrates that divergent evolution comes to halt when the residents come to lie on fitness maxima themselves, i.e., when they reach the values $\epsilon^*$ and $-\epsilon^*$ given by eq. (18). Parameter values: $s=4$, $\sigma_K=1$, $\sigma_\alpha=1.2$, $K_0=1$. 
}
\end{figure}

This can be made mathematically precise by considering the selection gradients in the two resident strains. For example, it is fairly easy to show that
\begin{align}
\left.\frac{\partial f(\epsilon,-\epsilon,y)}{\partial y}\right\vert_{y=\epsilon}=\frac{s\epsilon^{s-1}\left((\frac{2}{\sigma_\alpha})^s-\frac{2\left(1+\exp[2^{s-1}(\frac{\epsilon}{\sigma_\alpha})^s]\right)}{\sigma_K^s}\right)}{4\left(1+\exp[2^{s-1}(\frac{\epsilon}{\sigma_\alpha})^s]\right)}.
\end{align}
In particular, this selection gradient vanishes for $\epsilon^*$ satisfying  
\begin{align}
\left(\frac{2}{\sigma_\alpha}\right)^s=\frac{2\left(1+\exp[2^{s-1}(\frac{\epsilon^*}{\sigma_\alpha})^s]\right)}{\sigma_K^s},
\end{align}
an equation that can be solved to yield 
\begin{align}
\epsilon^*=2^{\frac{1-s}{s}}\sigma_\alpha\left\{\ln\left[2^{s-1}\left(\frac{\sigma_K}{\sigma_\alpha}\right)^s-1\right]\right\}^{\frac{1}{s}}.
\end{align}
Clearly, a solution $\epsilon^*>0$ exists if 
\begin{align}
\sigma_\alpha<2^{\frac{s-2}{s}}\sigma_K.
\end{align}
It follows that if inequality (19) is satisfied, there will be some resident pair $(\epsilon,-\epsilon)$ with $\epsilon$ small enough for which the selection gradient in the resident $\epsilon$ will be positive, and hence selection will drive this resident away from the singular point (since we assume $\epsilon>0$). By symmetry, the selection gradient in the resident $-\epsilon$ will be negative, so that overall, selection will drive evolutionary divergence of the two coexisting strains. Moreover, $(\epsilon^*, -\epsilon^*)$ is the equilibrium of the adaptive dynamics of the two coexisting and diverging strains, where $\epsilon^*$ is given by (18). At this equilibrium, both resident strains are maxima for the invasion fitness function (15), as illustrated in Figure 1.

Overall, assuming $s>2$ in the ecological functions (8) and (9), we now have the following situation. For $\sigma_K<\sigma_\alpha<2^{(s-2)/s}\sigma_K$, the singular point $x^*$ is a CSS, i.e., is convergent stable and evolutionarily stable, but two strains $\epsilon$ and $-\epsilon$ with $\epsilon>0$ small enough can mutually invade each other and hence coexist, and selection generates evolutionary divergence in such coexisting strains. Thus, for $\sigma_K<\sigma_\alpha<2^{(s-2)/s}\sigma_K$ the singular point is both a CSS and satisfies the conditions for evolutionary branching. Figure 2 summarizes these results.

\begin{figure}[htp]
\centering
\epsfig{file=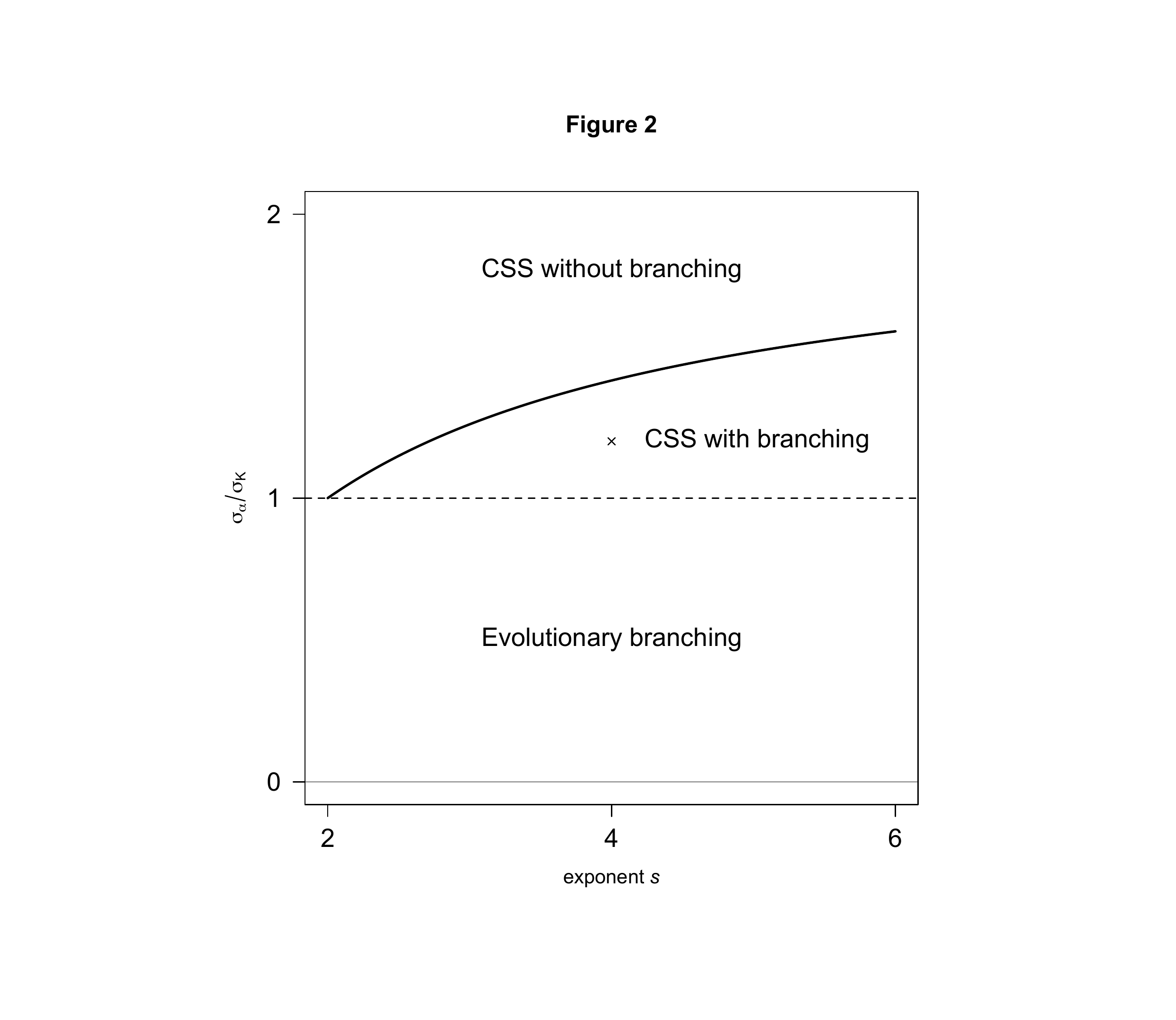,width=0.8\textwidth}
\caption{\label{Figure 2}
\footnotesize Parameter regions for evolutionary branching for adaptive dynamics based on the ecological functions (8) and (9). For $s>2$ there is a region of parameter space, bordered by the curves $\sigma_\alpha/\sigma_K=1$ and  $\sigma_\alpha/\sigma_K=s^{(s-2)/s}$, in which the adaptive dynamics (10) has a CSS at $x^*=0$ that can be the starting point of evolutionary branching. The point indicated by a cross corresponds to the parameter values used for Figures 3 and 4.
}
\end{figure}

However, there is a snag, which is that while phenotypes on either side of the singular may invade each other for $\sigma_K<\sigma_\alpha$, they cannot invade the singular strategy, which is, after all, evolutionarily stable under these conditions. As a consequence, the establishment of a coexisting coalition $(\epsilon,-\epsilon)$ cannot occur deterministically from very low frequencies. However, establishment of such coalitions can occur when the density of the mutants $\epsilon$ and $-\epsilon$ reaches a certain threshold. In fact, the ecological systems consisting of three strains $\epsilon$, $0$ (the singular strategy) and $-\epsilon$ exhibits bistability, as can be seen as follows. 

Given three strains $(\epsilon,0,-\epsilon)$, and assuming an idealized invasion scenario in which both $\epsilon$ and $-\epsilon$ have density $\delta$ and the singular strain has density $K(0)-2\delta=K_0-2\delta$, the per capita growth rate of the marginal strains is given by
% \begin{align}
% 1-\frac{\alpha(\epsilon,0)(K(0)-2\delta)+\delta(\alpha(\epsilon,\epsilon)+\alpha(\epsilon,-\epsilon))}{K(\epsilon)},
% \end{align} 
% while the per capita growth rate of the singular strain is
% \begin{align}
% 1-\frac{\alpha(0,0)(K(0)-2\delta)+\delta(\alpha(0,\epsilon)+\alpha(0,-\epsilon))}{K(0)}.
% \end{align}
\begin{align}
1-\frac{\alpha(\epsilon,0)(K_0-2\delta)+\delta(\alpha(\epsilon,\epsilon)+\alpha(\epsilon,-\epsilon))}{K(\epsilon)},
\end{align} 
while the per capita growth rate of the singular strain is
\begin{align}
1-\frac{\alpha(0,0)(K_0-2\delta)+\delta(\alpha(0,\epsilon)+\alpha(0,-\epsilon))}{K_0}.
\end{align}
Because the singular point is an ESS (always assuming $\sigma_K<\sigma_\alpha$), it follows that (21) is larger than (20) for $\delta$ (and $\epsilon$) small enough, so that the ecological equilibrium at which only the singular strain has positive density is locally stable. However, as $\delta$ becomes larger and reaches a certain threshold, the growth rate of the marginal strains becomes larger than the growth rate of the singular strains. By equating the two growth rates (20) and (21), one can find an analytical expression for the threshold value of $\delta^*$,
\begin{align}
\delta^*=K_0\frac{(\sigma_{\alpha}/\sigma_K)^s-1}{2^s-4} + {\cal{O}}(\epsilon^{s})
\end{align}
 and it is possible to show that for densities of the marginal strain above this threshold, the growth rate of the marginal strains is always larger than the growth rate of the singular strain. This implies that once the density of the marginal strains has crossed the threshold, the singular strain will go extinct, and coexistence of the coalition $(\epsilon,-\epsilon)$ will be established. Thus, the ecological equilibrium at which only the marginal strains have positive density is also locally stable.

In fact, this threshold value can be fairly small, in which case one would expect that it can be reached due to random fluctuations in models incorporating demographic stochasticity. Indeed, it is easy to observe evolutionary branching after convergence to a CSS in individual-based implementations of the competition models described here. In such models, individuals are characterized by their trait value, and their death rate is determined by the carrying capacity on the one hand, and by the effective density on the other hand. For an individual with trait value $x$, the effective density $N_{eff}$ is obtained by summing over all other individuals in the population, where each individual $y$ is counted with a weight $\alpha(x,y)$: $N_{eff}=\sum_y\alpha(x,y)$. The death rate of individual $x$ is then $N_{eff}/K(x)$, and all individual are assumed to have a birth rate of 1. Individual birth and death rates then define a stochastic evolutionary process, as described in detail in \cite{dieckmann_etal2004} (Chapter 4) and in \cite{doebeli2010}. When an individual gives birth during this process, the phenotype of the offspring is chosen from the mutation kernel, a normal distribution that has a mean equal to the parent value and some small variance $\sigma_{mut}$. Note that in such a model, all individuals present at any point in time have different phenotypes, and when the average phenotype of the population is close to the singular value, the population actually consists of a cluster of phenotypes around the singular value. Because birth and death events are stochastic, phenotypes at a distance $\geq\epsilon$ from the singular point may reach non-negligible densities due to random fluctuations even if they are not favored deterministically by selection. As a consequence of these stochastic fluctuations, evolutionary branching can occur after convergence to a CSS in accordance with the analytical arguments given above. This is illustrated in Figure 3.

\begin{figure}[htp]
\centering
\epsfig{file=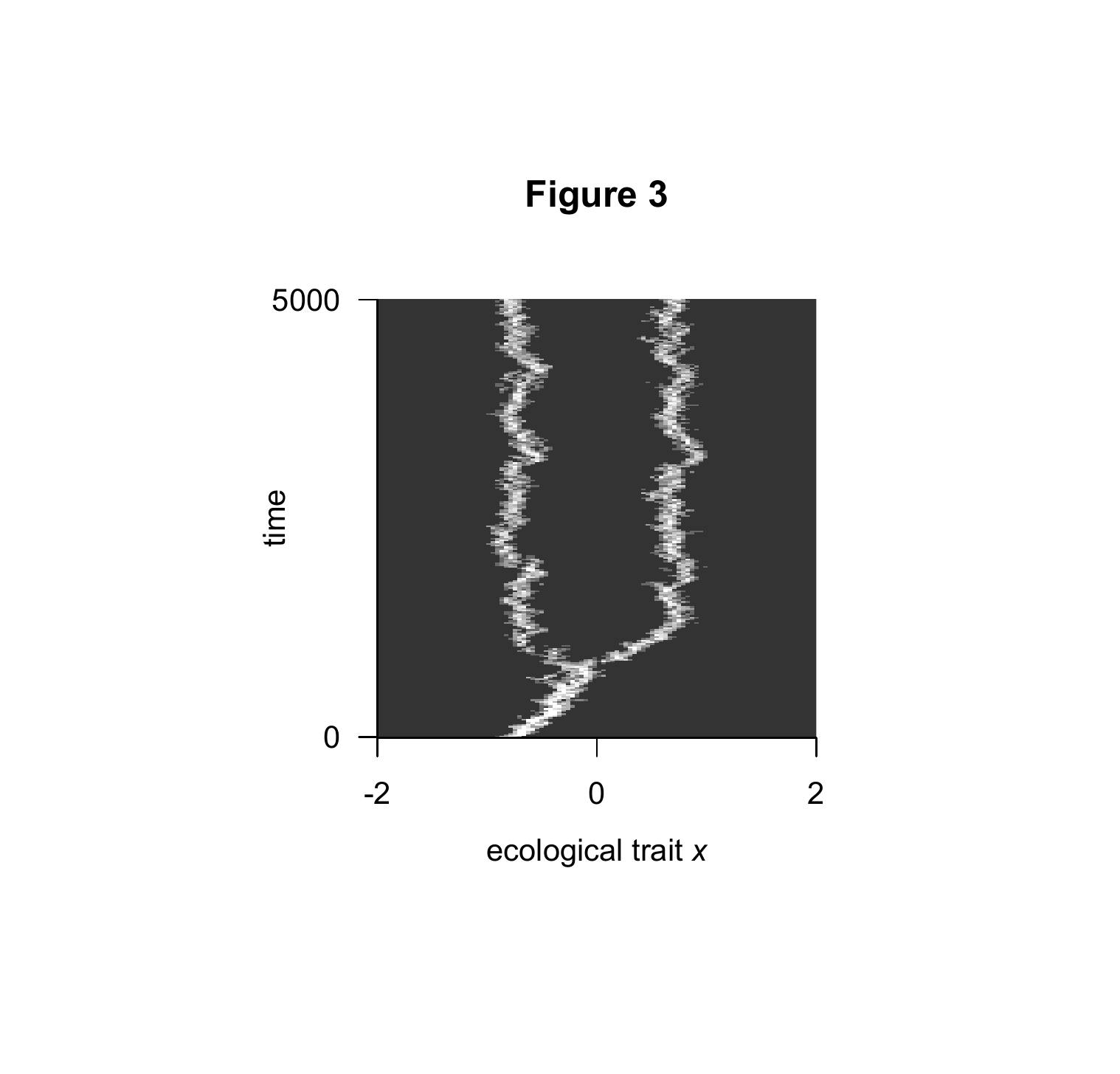,width=1\textwidth}
\caption{\label{Figure 3}
\footnotesize Evolutionary branching from a CSS in the individual-based model. Parameter values were $s=4$, $\sigma_K=1$, $\sigma_\alpha=1.2$, $K_0=270$, and $\sigma_{mut}=0.01$ for the width of the mutation kernel. The population was initialized with individuals drawn from a Gaussian distribution with mean $-1$ and a small variance.
}
\end{figure}

\newpage 
\vskip 2cm
{\bf \large Discussion}
\vskip 1cm

The theory of adaptive dynamics \citep{metz_etal1996, geritz_etal1998, dieckmann_law1996} has established itself as an excellent mathematical toolbox for understanding long-term evolutionary dynamics of quantitative traits. In particular, the paradigmatic phenomenon of evolutionary branching has been useful for understanding the ecological conditions leading to adaptive diversification and speciation \citep{dieckmann_doebeli1999, kisdi_geritz1999, dieckmann_etal2004, doebeli2010}. During the process of evolutionary branching, a population evolves under directional selection to a point in phenotype space where selection turns disruptive due to frequency-dependence, after which the population splits into diverging phenotypic clusters.

The point at which selection ceases to be directional is a so-called singular point, and if the trait under consideration is 1-dimensional, i.e., given by a real number, the conditions for evolutionary branching can be described by properties of the invasion fitness function at the singular point. The first derivative of the invasion fitness function vanishes by definition at a singular point, and if the second derivative is non-zero, then the singular point is an evolutionary branching point if and only if it is convergent stable and evolutionarily unstable \citep{geritz_etal1998, diekmann2003, doebeli2010}. These conditions already reveal that evolutionary branching is a generic phenomenon that occurs not only at particular points in parameter space, but for whole parameter regions. Here we showed that the realm of evolutionary branching can be even increased if the invasion fitness vanishes to higher than first order in the Taylor expansion around a singular point. More precisely, under these conditions a Continuously Stable Strategy (CSS), i.e., a convergent stable and evolutionarily stable singular point, can be the starting point for evolutionary branching, which is never the case when the second derivative of the invasion fitness function at the singular point is non-zero.

The mechanism for the observed phenomenon can be appreciated graphically by considering a perturbation of the invasion fitness function at the singular point (Figure 1), i.e., by comparing the two invasion fitness functions $F(z)=f(x^*-\epsilon,x^*+\epsilon,z)$ and $\hat F(z)=f(x^*,z)$, where $x^*$ is the singular point, $\epsilon>0$ is small, and $z$ is the phenotype of a rare mutant. If $x^*$ is a CSS, then $\hat F$ has a maximum at $x^*$. If $d\hat F(x^*)/dz\neq0$, then to second order, $\hat F$ is a parabola with maximum at $x^*$, and it follows that the perturbed function $F(z)$ is also a parabola with negative curvature, and with zeroes at $x^*-\epsilon$ and $x^*+\epsilon$. This fitness function therefore generates selection towards the singular point in both perturbed resident strains.

However, if $d\hat F(x^*)/dz=0$, then the multiplicity of the maximum at $x^*$ is higher than 1, and as a consequence, the perturbed function $F(z)$ has two maxima in the vicinity of $x^*$, one lying to the left of $x^*-\epsilon$ and the other lying to the right of $x^*+\epsilon$. Therefore, selection in the two perturbed residents is away from the singular point, as illustrated in Figure 1b. Thus, in terms of selection, when the second derivative of $\hat F$ at $x^*$ is zero resolution of the multiplicity of the maximum of $\hat F$ at $x^*$ can result in a scenario that is opposite to the situation occurring when the second derivative of $\hat F$ at $x^*$ is non-zero (and hence $x^*$ is a ``regular'' CSS).

We have described the parameter range for which this happens using a general class of models for frequency-dependent competition. In particular, CSS's, which are usually thought to be endpoints of the evolutionary dynamics, can be the starting point for evolutionary diversification due to frequency-dependent competition. Recently Frans Jacobs has made similar observations for invasion fitness functions that are given as fourth-order polynomials around the singular point \citep{jacobs2010}. In our models, the exponent of the first non-zero order in the Taylor expansion of the invasion fitness function around the singular is the same as the exponent $s$ appearing in the carrying capacity function (8) and in the competition kernel (9). We have assumed that $s\geq2$ to ensure twice differentiability of the invasion fitness at the singular point. For $s=2$, the model converts to the familiar Gaussian model, in which the singular point is a CSS for $\sigma_K<\sigma_\alpha$, and an evolutionary branching point for $\sigma_K>\sigma_\alpha$, where $\sigma_K$ and $\sigma_\alpha$ are the curvatures of the competition kernel and the carrying capacity. Because the second derivative of the invasion fitness function at the singular point is non-zero for $s=2$, no evolutionary diversification occurs in this case when the singular point is a CSS, i.e., when $\sigma_K<\sigma_\alpha$. For $s>2$, the singular point is still a CSS for $\sigma_K<\sigma_\alpha$, but because the second derivative at the singular point is 0, evolutionary branching can occur for $\sigma_K<\sigma_\alpha<2^{(s-2)/s}\sigma_K$, i.e., for a range of parameters for which the CSS condition holds. Note that for all $s$ evolutionary branching still occurs for $\sigma_K>\sigma_\alpha$. (In fact, one can show that for exponents $0<s<2$, the threshold $\sigma_K=\sigma_\alpha$ remains the boundary between evolutionary diversification on the one hand, and CSS's as endpoints of the evolutionary dynamics on the other hand.) Thus, compared to models in which the invasion fitness function only vanishes to first order at the singular point, the region of parameter space leading to evolutionary diversification is enlarged when the invasion fitness function also vanishes to second order. We note that this phenomenon also occurs when the exponents in the carrying capacity and the competition kernel are different. Based on both analytical and numerical studies, and denoting these exponents by $s_K$ and $s_\alpha$, respectively, we conjecture that branching can occur after converging to a CSS whenever $s_\alpha>2$ and $s_K\gtrsim s_\alpha$.

To conclude, we mention an interesting connection of the adaptive dynamics models studied here to the corresponding logistic partial differential equations models
\begin{align}
\frac{\partial \phi(x)}{\partial t}=r\phi(x)\left(1-\frac{\int \alpha(x,y)\phi(y)dy}{K(x)}\right),
\end{align} 
which describe the dynamics of phenotype distributions $\phi(x)$ due to frequency-dependent competition. Here $K(x)$ and $\alpha(x,y)$  are the carrying capacity function (8) and the competition kernel (9), and the integral $\int \alpha(x,y)\phi(y)dy$ is the effective density experienced by individuals of phenotype $x$. As before, the dynamics (23) can be studied for different exponents $s$ of the ecological functions (8) and (9). For the Gaussian case $s=2$, it is well known that system (23) has an equilibrium distribution consisting of a single delta peak at $x^*=0$ for $\sigma_K<\sigma_\alpha$, and a Gaussian equilibrium distribution with maximum at $x^*=0$ and a positive variance for $\sigma_K<\sigma_\alpha$. This corresponds to the results from adaptive dynamics in the sense that the boundary for $\sigma_K=\sigma_\alpha$ separates two different dynamic regimes, one in which diversity is not maintained and another in which it is, even though maintenance of diversity in this latter regime occurs in the form of unimodal phenotype distributions, rather than in the form of multimodal distributions representing distinct phenotypic clusters. \cite{pigolotti_etal2009} have shown that the Gaussian case $s=2$ is structurally unstable, and that for $s>2$ the regime in which diversity is maintained is characterized by multimodal equilibrium distributions. Here we can add that not only does model (23) have equilibrium distributions consisting of multiple delta peaks when $s>2$, but such diversification can occur even for $\sigma_K<\sigma_\alpha$, as illustrated in Figure 4. In fact, our numerical simulations indicate that diversification can occur in the partial differential equation model (23) exactly for the same range of parameters as in the adaptive dynamics model, i.e., for $\sigma_\alpha<2^{(s-2)/s}\sigma_K$. 

\begin{figure}[htp]
\centering
\epsfig{file=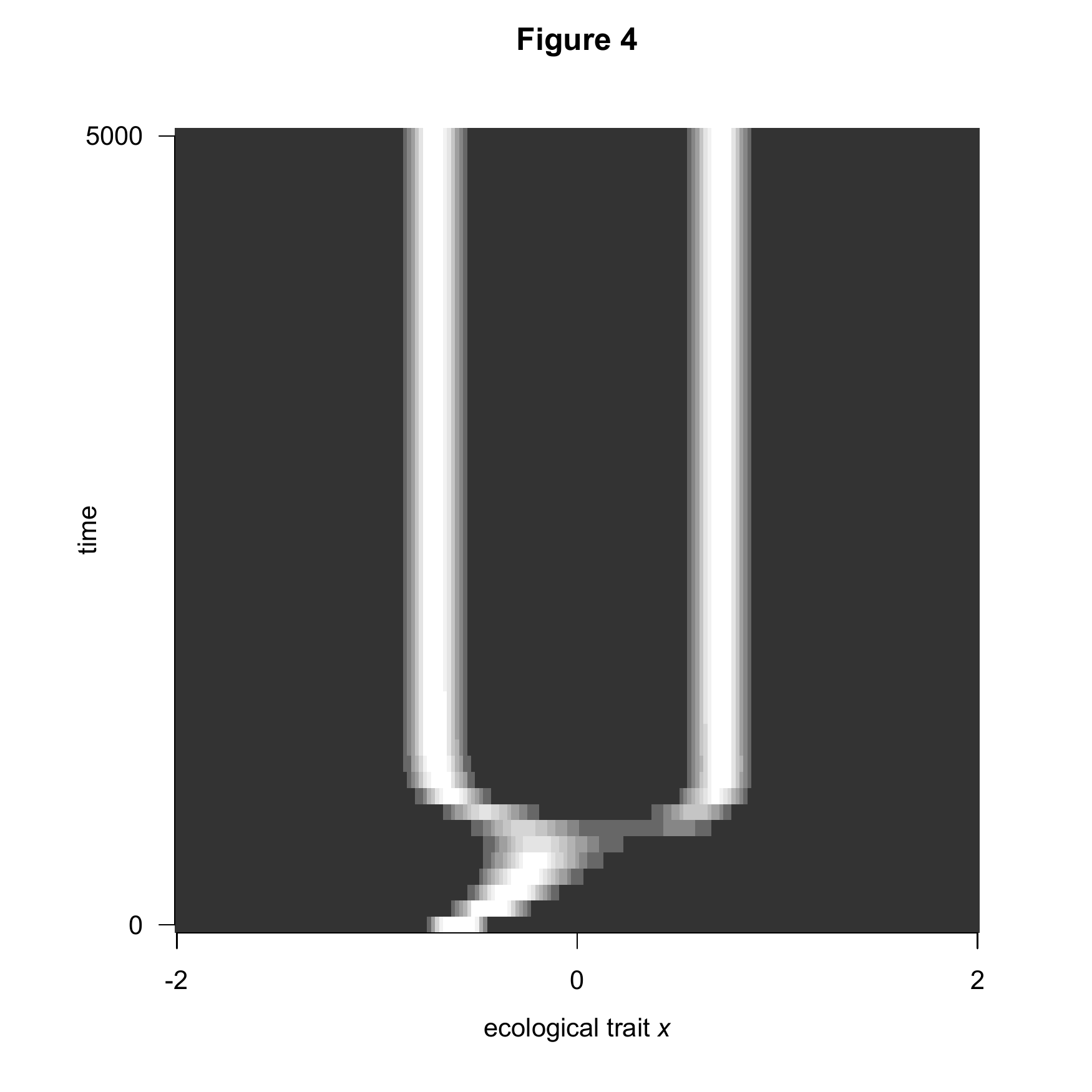,width=0.7\textwidth}
\caption{\label{Figure 4}
\footnotesize Evolutionary diversification in the form of a bimodal equilibrium distribution branching in the logistic partial differential equation model (23). Parameter values were the same as for Figure 3. A convolution with a Gaussian mutation kernel of width 0.01 was incorporated in the birth term in (23) to make the temporal evolution consistent with the individual-based simulation shown in Figure 3. Note that the position of the two modes at equilibrium corresponds well with the two phenotypic clusters resulting from evolutionary branching in Figure 3.
}
\end{figure}

Moreover, for $\sigma_K<\sigma_\alpha<2^{(s-2)/s}\sigma_K$, and in complete agreement with the adaptive dynamics model, the dynamics (23) converges to a single delta peak at $x=0$ if the initial phenotype distribution is very narrow, and diversification can only occur if the initial distribution is sufficiently wide. Overall, however, the partial differential equation models confirm that for exponents $s>2$ in the ecological functions, the region of parameter space leading to adaptive diversification is enlarged.

\newpage
\bibliography{EvolutionofDiversity}
\bibliographystyle{prslb}

\end{document}